\documentclass[aps,pra,preprint,groupedaddress]{revtex4-2}
\usepackage{graphicx}
\usepackage{dcolumn}
\usepackage{bm}
\usepackage{cancel}

\usepackage{color}
\usepackage{ulem}
\usepackage{xcolor}
\usepackage{physics} 
\definecolor{Cerulean}{rgb}{0.,0.59,0.835}
\definecolor{RubineRed}{rgb}{0.61,0.07,0.12}
\usepackage[colorlinks=true,citecolor=Cerulean,linkcolor=RubineRed,urlcolor=Cerulean,pdftex]{hyperref}

\input{epsf}

\begin{document}
\title{Nucleation of helium in liquid lithium} \author{J.Mart\'\i,
  F.Mazzanti, G.E.Astrakharchik, L.Batet}
\email{jordi.marti@upc.edu,ferran.mazzanti@upc.edu,grigori.astrakharchik@upc.edu,lluis.batet@upc.edu}
\affiliation{Department of Physics, Polytechnical University of
  Catalonia-Barcelona Tech, \\ B5-209 Northern Campus, Jordi Girona
  1-3, 08034 Barcelona, Catalonia, Spain.}

\author{L.Portos-Amill, B.Pedre\~{n}o}
\affiliation{Barcelona School of Telecommunications Engineering, \\
Polytechnical University of Catalonia-Barcelona Tech, B3 Northern Campus, \\
Jordi Girona 1-3, 08034 Barcelona, Catalonia, Spain.}

\date{\today}

\begin{abstract}
Fusion energy stands out as a promising alternative for a future decarbonised energy system. To be sustainable, future fusion nuclear reactors will have to produce their own tritium. In the so-called breeding blanket of a reactor, the neutron bombardment of lithium will produce the desired tritium, but also helium, which can trigger nucleation mechanisms owing to the very low solubility of helium in liquid metals. An understanding of the underlying microscopic processes is important for improving the efficiency, sustainability and reliability of the fusion energy conversion process. A spontaneous creation of helium drops or bubbles in the liquid metal used as breeding material in some designs may be a serious issue for the performance of the breeding blankets. This phenomenon has yet to be fully studied and understood. This work aims to provide some insight on the behavior of lithium and helium mixtures at experimentally corresponding operating conditions (843~K and pressures between 0.1 and 7~GPa). We report a microscopic study of the thermodynamic, structural and dynamical properties of lithium-helium mixtures, as a first step to the simulation of the environment in a nuclear fusion power plant. We introduce a microscopic model devised to describe the formation of helium drops in the thermodynamic range considered. A transition from a miscible homogeneous mixture to a phase-separated one, in which helium drops are nucleated, is observed as the pressure is increased above 0.175~GPa.  The diffusion coefficient of lithium (2~\AA$^2$/ps) is in excellent agreement with reference experimental
data, whereas the diffusion coefficient of helium is in the range of 1~\AA$^2$/ps and tends to decrease as pressure increases.  The radii of helium drops have been found to be between 1 and 2~\AA.
\end{abstract}

\maketitle

\section{Introduction}
\label{intro}

Within the framework of future energy supply, with the constraints posed by the need of electrification 
of the final energy demand, and the quest for more sustainable power generation methods in order to 
achieve a decarbonised electricity system, nuclear fusion energy stands out as a promising alternative.  
The fusion reaction that results most convenient in the present state of technological development is:

\begin{equation} 
{\rm D} + {\rm T} \rightarrow \;^{4}{\rm He} + {\rm n}+ 17.6\; {\rm MeV},
\label{eq0}
\end{equation} 
where 'D' stands for deuterium, 'T' for tritium and 'n' for a free neutron and where helium is a 
by-product\cite{kordavc2017helium}.  Deuterium is abundant in water, but tritium ($t_{1/2}$=12.3 year) 
must be artificially created.  Therefore,  in order for fusion energy to be sustainable, it is necessary that 
tritium be produced in the reactor itself. Tritium will be generated by means of the reactions of 
neutrons escaping from the plasma with lithium in the so-called breeding blankets (see for
 instance  \cite{federici2019overview} for an overview of these relevant components in DEMO, 
a demonstration power plant contemplated in the European Roadmap to Fusion). Breeding 
blankets (BB) will perform two additional functions besides producing tritium: extraction of fusion heat,
and shielding the magnets (superconducting coils) from the radiation escaping the plasma.  

Lithium has two natural isotopes $^6$Li (abundance 7.5 $\%$) and $^7$Li (92.5 $\%$), both 
producing tritium when capturing a neutron\cite{rubel2019fusion}:

\begin{equation} 
{\rm n} + \;^{6}{\rm Li} \rightarrow  {\rm T} + \;^{4}{\rm He} +  4.78\;{\rm MeV}
  \label{eq0-1}
\end{equation}
\begin{equation} 
{\rm n} + \;^{7}{\rm Li} \rightarrow  {\rm T} + \;^{4}{\rm He} 
+ {\rm n} -  2.47\;{\rm MeV}
\label{eq0-2}
\end{equation} 
Tritium self-sufficiency will require a certain neutron multiplication in order to close the fuel 
cycle with a net gain so that the so-called tritium breeding ratio is greater than 1.  In order to 
fulfill their functions, some BB designs feature solid (ceramic) breeders cooled by helium, 
while others rely on a liquid metal (LM) cooled by helium or water.  Usually the components 
of BB designs are lithium-lead eutectic (LLE)\cite{coen1985lithium,de2008lead,federici2019overview}.
Besides $^7$Li,  lead will provide some fast neutron multiplication (neutrons hit the walls 
of the reaction chamber with energies bigger than 14 MeV).  As shown in Eqs.(\ref{eq0-1}) 
and (\ref{eq0-2}),  He is produced mol-to-mol along with T.  However,  He is practically 
insoluble in the liquid metal (Henry's constant for helium in Li at 843 K would be 
around 7$\times 10^{-14}$ Pa$^{-1}$ atomic fraction; for LLE it is estimated to be 
lower\cite{sedano2007helium}). Tritium self-sufficiency requirement is thus linked to 
a possible super-saturation of helium in the liquid metal and, consequently, to a 
possible nucleation of helium in the form of bubbles.  This phenomenon may have 
a great impact in the performance of the BB: changes in the magnetohydrodynamic 
flow, affectation of the heat transfer, and changes in the tritium migration 
mechanisms.  Other systems that could be affected by helium nucleation are, for 
instance, free-surface Li first wall concepts\cite{ruzic2017flowing,smolentsev2019integrated} 
and the Li jet targets in the future International Fusion Materials Irradiation 
Facility\cite{nakamura2008latest}.

\begin{figure}[htbp]
\begin{center}
  \includegraphics[width=1.2\columnwidth]{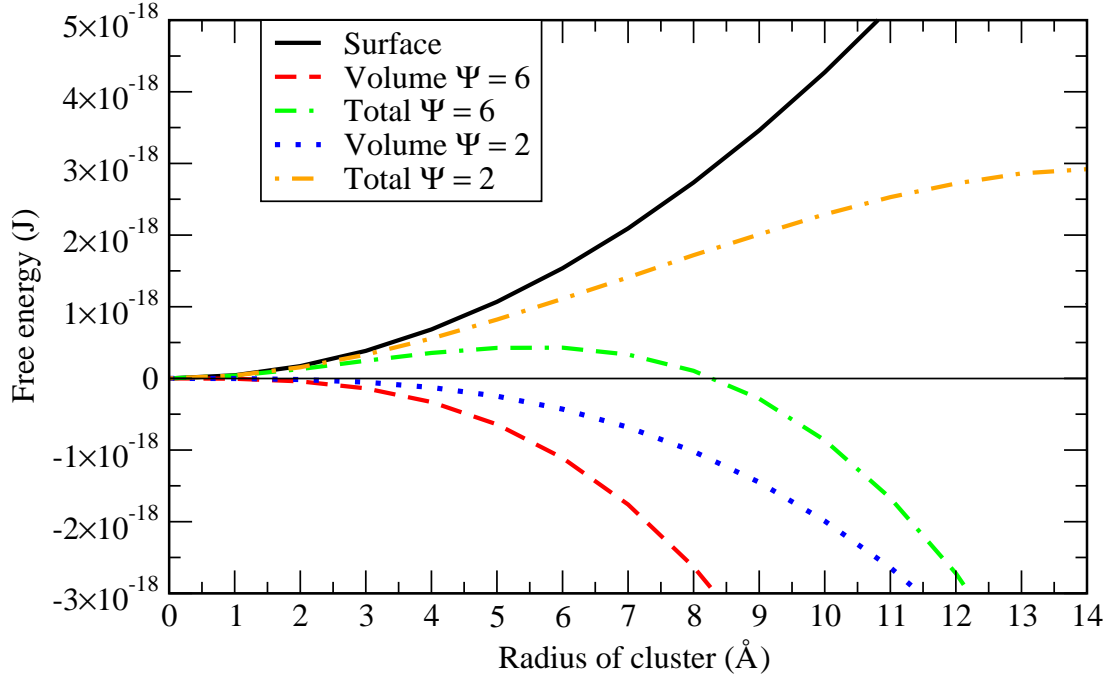}
\caption{\label{fig1}Free energy of a cluster of He forming in lithium at 843 K assuming a 
surface tension of 0.34 N/m \cite{yakimovich2000experimental}. Using a volume of 
17~\AA$^3$ for helium\cite{batet2011numeric} critical size is 42 atoms when supersaturation 
ratio $\Psi = 6$ and 110 atoms when $\Psi = 2$. }
\end{center}
\end{figure}

In the quest for tools to model the effect of the undesired helium bubbles being formed
in the blanket walls of a nuclear fusion plant, helium nucleation models must be developed.  
So far, no experiment exists allowing to validate such models. The low solubility of 
He in LM makes computer simulations extremely expensive if trying to capture the onset 
of nucleation at the design operational pressures and temperatures of BB.  To provide 
an order of magnitude, a rough estimation based in the Gibbs' Classical Nucleation Theory
\cite{gibbs1878equilibrium, gibbs1906scientific} (CNT) follows from Ref. \cite{batet2011numeric}.  
At a temperature of 843 K some 40 atoms of He are needed to form a stable cluster when 
He concentration is 6 times larger than its solubility (see Fig.~\ref{fig1}).  A simulation involving 
40 atoms of lithium at 843 K and 1 bar would require almost 10$^9$ atoms of lithium to be in 
those conditions.  If Henry's law is valid at high pressures, only 1000 atoms of lithium would 
be necessary at 100 GPa. 

This work focuses on the simulation of He-Li mixtures at high pressures as a first step towards 
the simulation of the Li-Pb-He mixtures at low pressure.  The goal is to capture the onset of 
the nucleation in order to advance towards the modeling of the phenomenon.  In this work 
we describe a mixture of helium and lithium atoms in the bulk, developing a microscopic 
model that is able to reproduce the helium-lithium mixture instability towards nucleation 
of helium drops, and performing ab-initio simulations of occurring processes.  
The Li-Li, He-He and He-Li pair interactions are fed as an input to both classical 
Monte Carlo (MC) and molecular dynamics (MD) simulations.  We find 
thermodynamic, structural and dynamic properties of lithium and helium mixtures 
at high temperatures and in a wide range of pressures.  Both MC 
and MD computational techniques have been previously proven to provide reliable
predictions for a wide variety of classical and quantum atomic and molecular systems, 
ranging from pure quantum systems including hydrogen and
helium\cite{mazzanti2004high,mazzanti2008ground,
pierleoni2016liquid,bombin2017dipolar,mazzola2018phase,sanchez2020supersolid} to 
classical molecular liquids in 
solution\cite{marti2002microscopic,nagy2007liquid,videla2011aqueous,sala2012specific,calero20151h}
and at interfaces\cite{gordillo2000hydrogen,zambrano2009thermophoretic,rodriguez2017surface}
and to highly complex biosystems such as proteins or 
membranes\cite{karplus1990molecular,karplus2002molecular,marti2004transition,yang2014diffusion}. 
We calculate and report thermodynamic properties such as the average internal energy as 
a function of pressure.  In order to quantify the spatial and dynamical 
structure we calculate the pair distribution functions, the mean squared displacements, 
and the velocity autocorrelation functions. We also obtain the diffusion coefficient of
lithium and helium at different pressures ranging from 0.1 to 10 GPa as well as the 
spectral densities of He and Li, reporting information on their main vibrational modes.

\section{METHODS}
\label{meth}

\subsection{Microscopic model}
\label{ffs}

We base our simulations on a microscopic model Hamiltonian describing a mixture of
$N_{\rm Li}$ lithium and $N_{\rm He}$ helium atoms,  which are taken to be point-like 
particles of mass $m_{\rm Li}$ and $m_{\rm He}$, respectively.   In order to reproduce the 
experimental conditions, we only consider situations where $N_{\rm Li} \gg N_{\rm He}$.
Each species is characterized by particle coordinates and velocities 
$\{ {\bf r}_{{\rm Li},i}, {\bf v}_{{\rm Li},i} \}$ and
$\{ {\bf r}_{{\rm He},j}, {\bf v}_{{\rm He},j} \}$,  with $i$ and $j$ spanning the ranges
$1,\ldots, N_{\rm Li}$ and $1,\ldots, N_{\rm He}$, respectively.  The Hamiltonian
of the system is then written as

\begin{eqnarray}
\label{eq1}
H & = & \frac{1}{2}\sum\limits_{i=1}^{N_{\rm Li}} m_{\rm Li}v_{{\rm Li}, i}^2
+\frac{1}{2}\sum\limits_{i=1}^{N_{\rm He}}m_{\rm He}v_{{\rm He}, i}^2\\
&+&\sum\limits_{i<j}^{N_{\rm Li}}V_{{\rm Li}-{\rm Li}}(|{\bf r}_{{\rm Li},i}-
{\bf r}_{{\rm Li},j}|)
+\sum\limits_{i<j}^{N_{\rm He}}V_{{\rm He}-{\rm He}}(|{\bf r}_{{\rm He},i}-
{\bf r}_{{\rm He},j}|)
+\sum\limits_{i=1}^{N_{\rm Li}}
\sum\limits_{j=1}^{N_{\rm He}}
V_{{\rm Li}-{\rm He}}(|{\bf r}_{{\rm Li},i}-{\bf r}_{{\rm He},j}|)\;,
\nonumber
\end{eqnarray}
where the first two terms describe the kinetic energy, while the last three terms account 
for the intra- and inter-species interaction, respectively.  Periodic boundary conditions are 
applied in order to minimize the finite-size effects and approximate better the properties of a large
system.  The typical simulation cage is a square box of length around 29$\;$\AA$\;$ for 
the reference pressure of 1 GPa. In lower pressure setups, box lengths are larger than 40~\AA.

\begin{figure}[htbp]
\begin{center}
  \includegraphics[width=1\columnwidth]{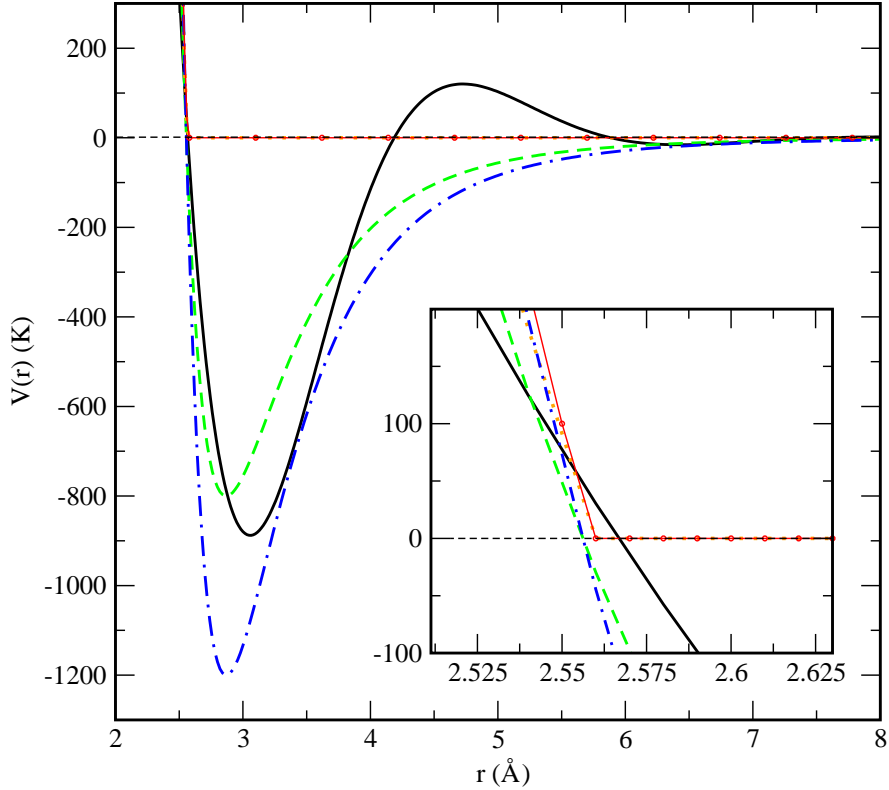}
\caption{\label{fig2}Microscopic two-body interaction potentials
  employed in this work.  Main figure: overall view of the interaction
  potentials $V(r)$ for Li-He, He-He and Li-Li as given
  by Eqs.~(\ref{eq2}-\ref{eq3}).  Li-Li (black line); He-He model
  1 (dot-dashed blue line); He-He model 2 (dashed green line); Li-He
  model 1 (red circles) and Li-He model 2 (dotted orange lines). The
  inset represents a zoom of the ``hard-wall'' area at short distances. }
\end{center}
\end{figure}

A crucial point of our model is an appropriate choice of the pair interaction potentials.  For
lithium-lithium interactions~(\ref{eq2}) we rely on the model proposed by Canales et
al. in Ref.~\cite{canales1993molecular,canales1994computer}, whereas the remaining interactions 
are a novelty of the present work.  The Li-Li pair interaction potential is modeled
as~\cite{canales1993molecular,canales1994computer}

\begin{equation}
V_{{\rm Li}-{\rm Li}}(r) = Ar^{-12} + B\exp{Cr}\cdot\cos{D(r-E)},
\label{eq2}
\end{equation}
where $r$ is the distance between the two atoms in Angstr\"{o}ms,  $V(r)$
is measured in Kelvin units and the potential coefficients are $A = 2.22125\times 
10^{7}\;$K\AA$\,^{12}$, $B = 41828.9$ K, $C = -1.20145$ \AA$^{-1}$, 
$D = 1.84959$ \AA$^{-1}$, $E = 5.03762\;$\AA. The interaction potentials are 
shown in Fig.~\ref{fig2} and they feature strong short-distance repulsion caused by 
Pauli exclusion due to overlapping electron orbitals, and a highly non-monotonic 
behavior which eventually follows a van-der-Waals attractive tail at large distances.  
The characteristic length of the interaction potential corresponds to the smallest 
distance at which the interaction changes sign,
$V_{\rm Li-Li}(\sigma_{\rm Li-Li}) = 0$, and is equal to 
$\sigma_{\rm Li-Li} = 2.5668\;$\AA.  
The characteristic energy scale is defined by the depth of the first minimum, 
equal to $\epsilon_{\rm Li-Li} \equiv V_{\rm Li-Li}(3.06) = -887.9$~K.

The helium-helium interaction has been considered to be of the
Lennard-Jones (LJ) type, and it has been parameterized to accurately
describe the system at the temperatures and pressures of interest 
which are well beyond ambient conditions. From preliminary simulations, 
we have found that the Aziz~II interaction potential{\cite{aziz1987new}, which 
is known to provide an excellent description of supefluid liquid helium at temperatures
close to absolute zero and moderate pressures around saturation density, 
does not work appropriately at the temperatures as high as 843~K and pressures 
in the GPa regime considered in the present study.  Instead, we retain the same 
width $\sigma_{\rm He-He} = 2.556\;$\AA\, but treating 
the potential depth $\epsilon_{\rm He-He}$ as a free adjustable parameter.  In this 
work we have found that, in order to be able to reproduce the nucleation process,
the depth must be increased to the typical values of the Li-Li potential~(\ref{eq2}). 
In order to test the influence of this interaction parameter, we considered two 
different values of the potential depth,  $\epsilon_{\rm He-He} = -1200$~K and 
$\epsilon_{\rm He-He} = -800$~K,  referred to as ``model~1'' and ``model~2'', respectively.

Finally the helium-lithium interactions have been modeled through a Lennard-Jones
potential at short distances, namely a ``hard" wall, with characteristic parameters 
given by the Lorentz-Berthelot rules obtained from the corresponding Li-Li and He-He values,
and a cutoff beyond that point. This results in $\sigma_{\rm Li-He} = 2.5615\;$\AA\, 
and $\epsilon_{\rm Li-He} = -1032.2\;$ K. The potential model is given by:

\begin{eqnarray}
V_{{\rm Li}-{\rm He}}(r) &=& 4\epsilon_{{\rm Li}-{\rm He}} \left[
  \left(\frac{\sigma}{r}\right)^{12} - \left(\frac{\sigma}{r}\right)^6
  \right], \quad r \le \sigma_{{\rm Li}-{\rm He}} \\ \nonumber &=& 0, \quad r >
\sigma_{{\rm Li}-{\rm He}}
\label{eq3}
\end{eqnarray}
The four considered pairwise interactions are shown in Fig.~\ref{fig2}.  In a very 
recent work \cite{zhou2021enabling}, it has been reported that specific
interatomic potentials based on Daw-Baskes and Finnis-Sinclair formalisms 
are able to describe the formation of helium bubbles in a palladium tritide 
lattice at temperatures of the order of 400 K and pressures in the range of 0.1 
to 2.2 GPa.  Furthermore, the formation of helium bubbles in tungsten
was also reproduced using purely repulsive He-W interaction potentials
in cluster simulations\cite{juslin2013interatomic}, models rather close to the 
ones presented in this work.

\subsection{Monte Carlo and molecular dynamics methods}

We rely on MC and MD methods to perform a series of computer simulations of the 
system. Both methods use the microscopic model as reported in Eq.}~(\ref{eq1}) 
to describe interactions between the atoms as an input and allow for the calculation 
of the thermodynamic quantities of interest.

The Monte Carlo method has been used to obtain the equilibrium properties
at fixed pressure $P$,  particle number $N$ and temperature $T$.  This
is done starting from the microscopic Hamiltonian of Eq.~(\ref{eq1}) and
sampling from it the corresponding Maxwell-Boltzmann distribution,  where the
probability of a state with energy $E$ is $p = \exp(-E/k_BT)$,  using the 
standard Metropolis algorithm.  Once the system has thermalized, we perform 
simulations to estimate properties such as the energy per particle and volume, 
as well as correlation functions such as the pair distribution function 
and the low-momentum static structure factor.  An advantage of the MC method 
is that it only uses the particle positions,  in contrast to MD where their 
velocities have also to be sampled.  This halves the number of microscopic 
variables to estimate, thus reducing the phase space and making the exploration 
very efficient.  This, however, comes at a price, since Monte Carlo can only sample equilibrium
configurations, and therefore it is not able to provide information about the time-dependent 
properties, in contrast to MD where the simulation propagates in real time.

In molecular dynamics, the force fields are also obtained from the model in Eq.~(\ref{eq1})
and the corresponding Newton's equations of motion, which are integrated numerically
using a standard leap-frog Verlet procedure\cite{frenkel2001understanding}.  In every 
simulation,  we have considered a fixed number of particles $N$ and pressure, $P$ while the 
volume is adjusted correspondingly.  In addition to the energetics and structural properties 
obtained also in MC,  MD provides access to time-dependent quantities such as the diffusion 
coefficient,  velocity autocorrelation functions and spectral densities.  As a stringent test of 
self-consistency, strict agreement between the common quantities sampled in MC and 
MD has to be obtained,  which requires the proper thermalization and averaging in both 
methods.

\section{RESULTS}
\label{res}

\subsection{Thermodynamic states}
\label{ts}

In all cases a homogeneous mixture of helium and lithium has been considered as the starting 
point of the simulations.  The concentration of helium has been set to $\sim$0.04
for a total of 40 helium atoms dissolved in a sea of 960 lithium atoms.  The main 
results for the thermodynamic quantities of interest obtained in both MC and 
MD are summarized in Table~\ref{tab1}.  Additional simulations at intermediate pressures 
(0.3 and 0.4~GPa for instance) have also been considered in several other sections of the 
manuscript.

\begin{figure}[htbp]
\begin{center}
\includegraphics[width=1\columnwidth]{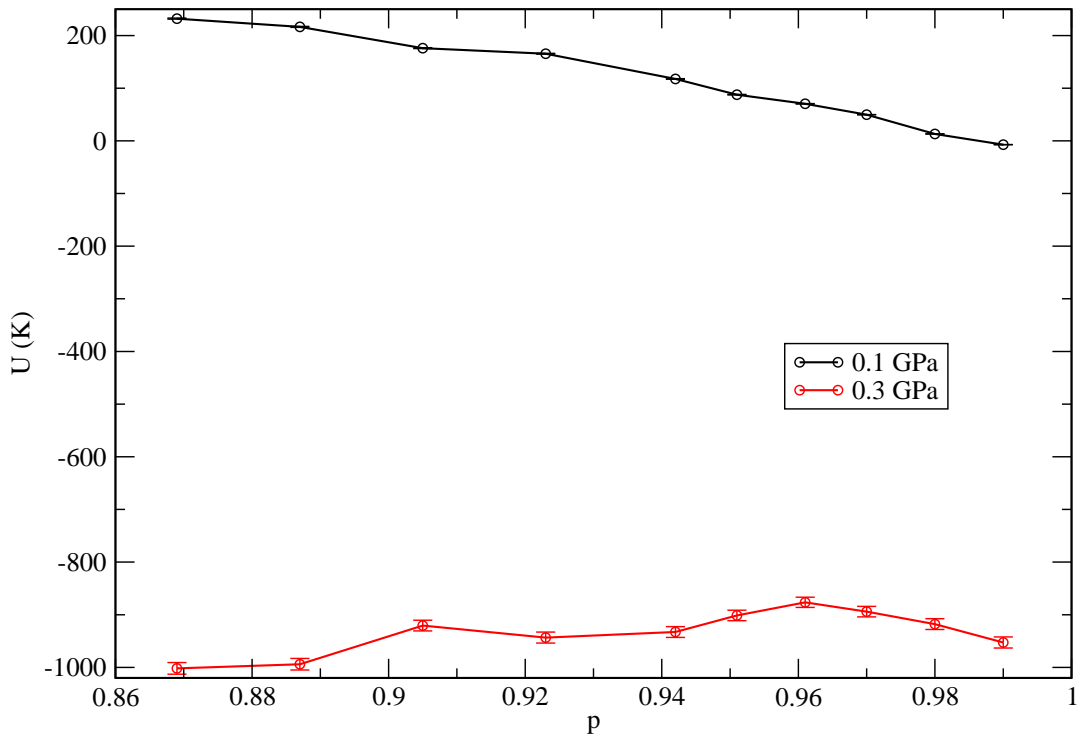}
\caption{\label{fig3} Total internal energies $U$ as a function of concentration $p$ for two characteristic pressures (0.1-0.3~GPa).}
\end{center}
\end{figure}

As a starting point of our analysis, we have obtained average internal energies and 
pressures of all considered simulation runs, as reported in Table~\ref{tab1}.  
In the MC simulations, the system was initially allowed to equilibrate
for a total of $10^7$ MC random movements and spanning a total time of about 50~ps
for the MD simulations. After that and in order to collect statistics the system was 
allowed to evolve for another $10^8$ 
random steps during the MC simulations. Correspondingly,  we collected MD trajectories 
200 ps long in all cases.  In both MC and MD statistical errors were less than 1$\%$ in all 
reported quantities. The state of lowest internal 
energy is found at the pressure of 1~GPa.  As an additional test, in order to explore the 
influence of helium concentration on the total energy of the system, we report energies 
as a function of the relative helium concentrations in Figure~\ref{fig3}. We defined the 
relative helium concentration $p$ as 

\begin{equation}
p = \frac{n_{\rm Li} - n_{\rm He}}{n_{\rm Li} + n_{\rm He}}
\label{eq3}
\end{equation}
We observe a monotonic behaviour at the lowest pressure (0.1 GPa),  that becomes non-monotonic  
for the second pressure (0.3 GPa). This might be an indication of a different qualitative phase coexistence 
for the two selected pressures.  We report further information about this aspect in the following sections.

\subsection{Pair distribution functions}
\label{rdf}

\begin{figure}[htbp]
\begin{center}
\includegraphics[width=1.2\columnwidth]{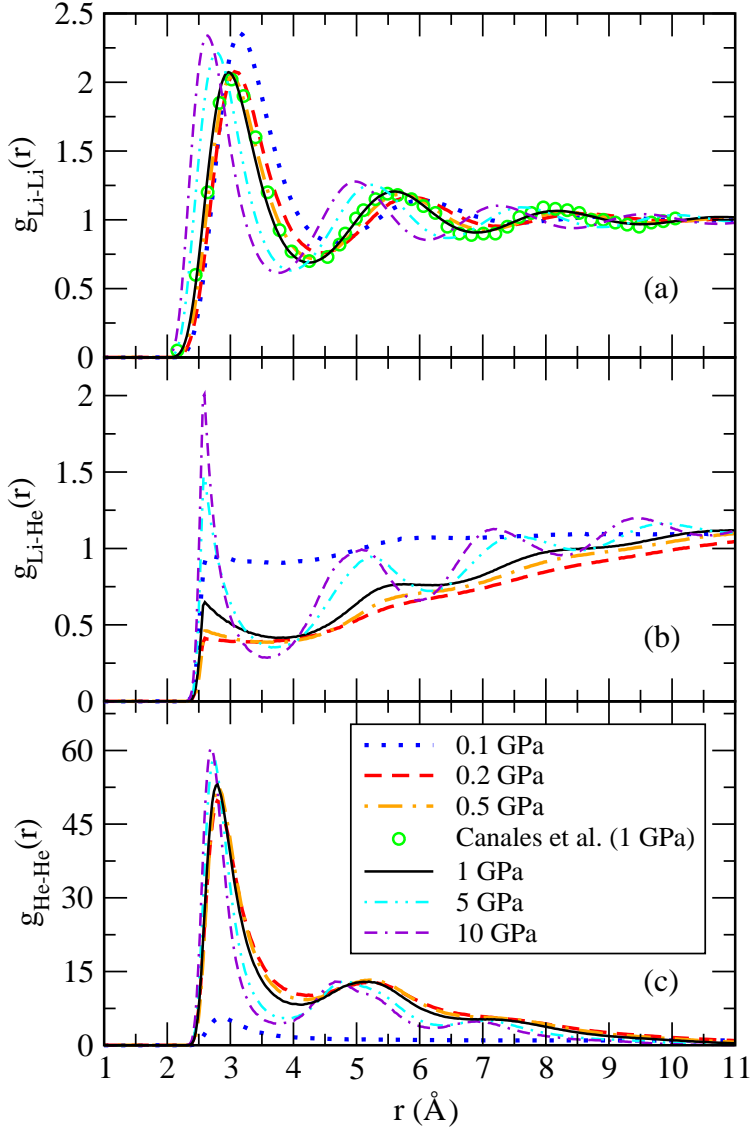}
\caption{\label{fig4} Pair distribution functions in a wide range of
  pressures (0.1-10~GPa) quantifying (a) He-He (b) He-Li (c) Li-Li
  correlations.  Green circles,  single-species Li-Li data from 
Ref.~\cite{canales1994computer}.  Lines: 0.1 GPa (dotted blue); 
0.2 GPa (dashed red); 0.5 GPa (dot-dashed orange);  
5 GPa (dot-dot-dashed cyan); 10 GPa (dash-dash-dotted violet).}
\end{center}
\end{figure}

In order to quantify the spatial correlations and visualize the structure of helium drops, 
we calculated the pair distribution function (RDF) in the simulated mixture of 960 Li and 
40 He atoms at 843~K.  The RDF is defined as

\begin{equation}
  g_{\alpha, \beta}(r) = {1 \over N_\alpha N_\beta}
  \sum_{i=1}^{N_\alpha} \sum_{j=1}^{N_\beta}
  \left\langle \delta( |{\bf r}|_{ij} - r)
  \right\rangle
\label{eq4}
\end{equation}
where $\alpha, \beta$ identify the species and the $\langle \cdots\rangle$ denotes a 
thermal average.  Being a two-particle correlator, it can measure translationally invariant 
ordering,  suitable to identify drop formation independently of its center of mass position.
Typical RDF for Li-Li, Li-He and He-He pairs are shown in Fig.~\ref{fig4} for different 
pressures.  The shape of the Li-Li pair distribution functions is characteristic of a liquid at 
equilibrium.  Li-Li RDF are hardly affected by the presence of a small concentration 
of helium atoms,  as can be seen in comparison with the behaviour of pure lithium at
1~GPa and the same temperature, taken from Ref.~\cite{canales1994computer}.   
One might also note that a change by a factor of 100 in the pressure does not significantly 
change the overall shape of $g_{\rm Li-Li}(r)$.  The short-range region is voided due
to the steep potential cores.  Strong oscillations are visible at separations
comparable to the mean interparticle distance, witnessing strong correlations in 
the liquid,  inducing shell effects.  At large distances, the pair distribution
function approaches a constant value,  thus confirming that lithium atoms are 
homogeneously filling the whole space.  The situation is drastically different
in the He-He RDF,  as they vanish at large distances as seen in Fig.~\ref{fig4}(c).
While at low pressure, $g_{\rm He-He}$ still shows a long-range plateau, this is not the 
case for large pressure where the RDF strongly decreases.  This implies that
helium atoms concentrate close to each other, thus forming droplets.  In this way, 
helium atoms form a miscible mixture on a lithium background at low pressure,
but have a tendency to phase separate at large pressures, splitting the
system into pure lithium and helium phases.  This scenario is further
supported by the huge increase in the height of the first and subsequent shells in a
He drop.  The drop size can be roughly estimated as the difference between
the distance at which $g_{\rm He-He}(r)$ significantly decays (position 
of the first minimum) and the position of the starting non-zero value of the RDF.

In order to verify the robustness of our analysis, we have compared the
results obtained with the two different He-He potential models proposed
corresponding to a depth well of 800\,K and 1200\,K, to find only minor changes.  
We have observed that when this depth is above 650~K, helium drops are not 
well formed and become significantly unstable in short time intervals of 
the order of 1~ps.  From here on,  the reported results are those corresponding 
to model ~1, as it predicts more stable helium droplets.

\begin{figure}[htbp]
\begin{center}
\includegraphics[width=1\columnwidth]{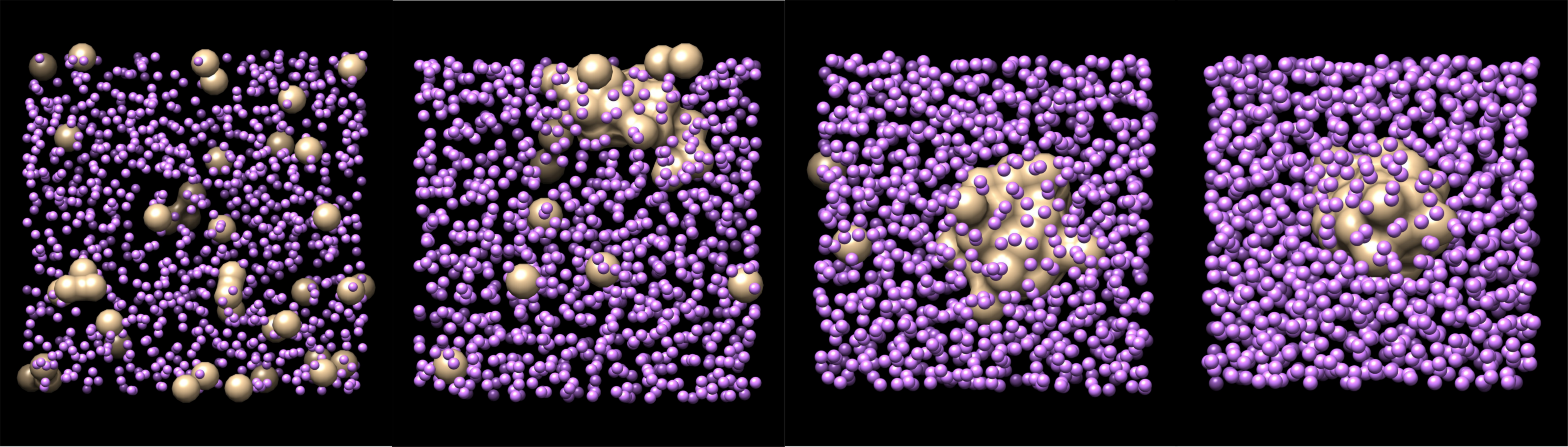}
\caption{\label{fig5} Snapshots of the He-Li mixtures at
  characteristic pressures: 0.1~GPa, 0.2~GPa, 1~GPa, 10~GPa
  (increasing pressure from left to right).}
\end{center}
\end{figure}

\begin{figure}[htbp]
\begin{center}
\includegraphics[width=1.\columnwidth]{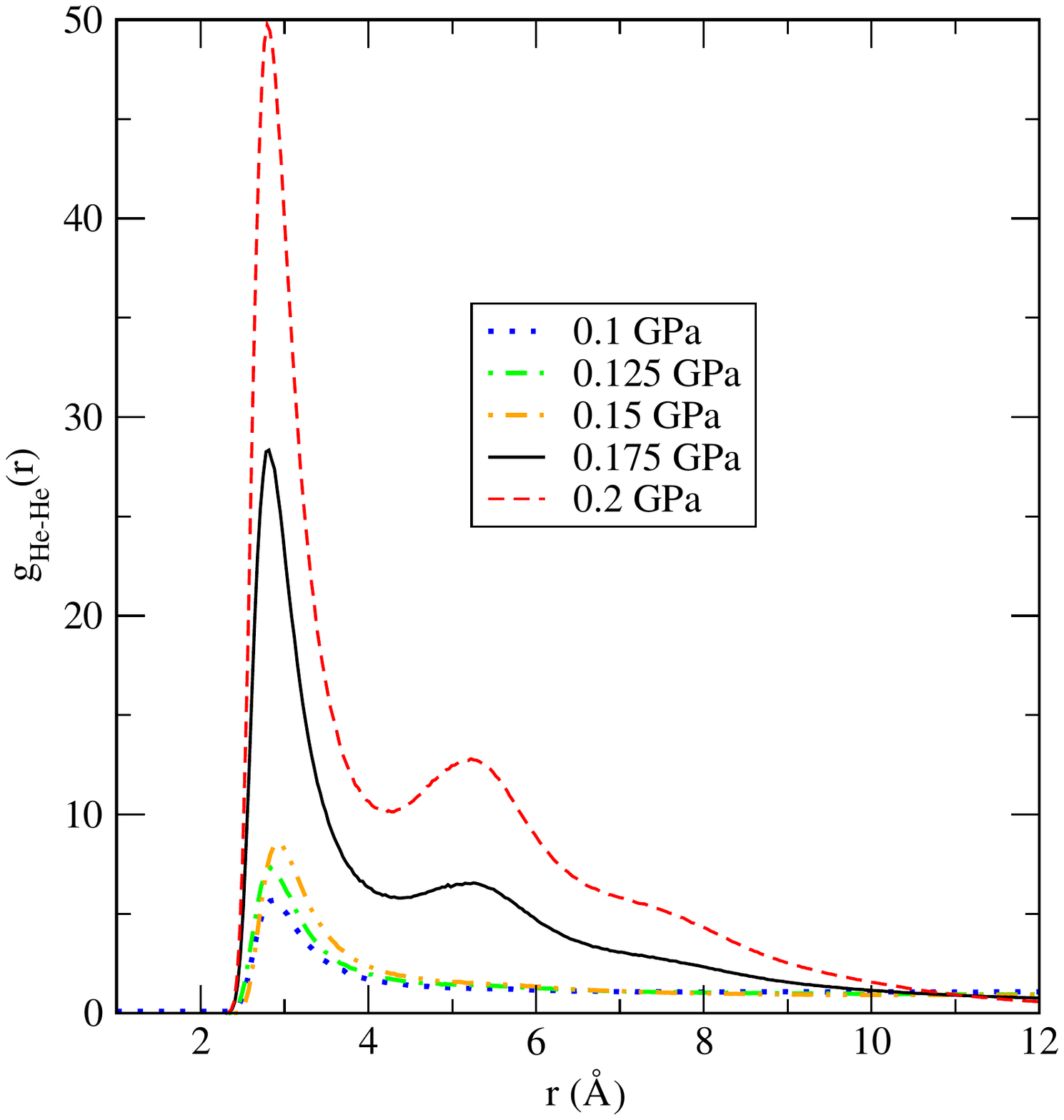}
\caption{\label{fig6} He-He pair distribution functions in the vicinity of the 
phase-separation transition (0.1-0.2 GPa).  Lines: 0.1 GPa (dotted blue); 
0.125 GPa (dot-dashed green); 0.15 GPa (dot-dot-dashed orange); 
0.175 GPa (black); 0.2 GPa (dashed red).}
\end{center}
\end{figure}

A set of four characteristic snapshots of the system at pressures
$P= 0.1, 0.2, 1$ and 10 GPa is shown in Fig.~\ref{fig5} to illustrate the
tendency of the system to form helium droplets when $P$ is increased above 
approximately 0.2~GPa.  At lower pressures helium is fully diluted in the
lithium bath. This is also seen in the He-He pair distribution function, which is 
shown in Fig.~\ref{fig6} for several values of $P$ close to the critical transition 
pressure.  There we can observe that helium drops start to appear at a crossover 
pressure around 0.175 GPa, which corresponds to phase separation 
(helium droplets in liquid lithium),  while full stable ones showing up at 0.2~GPa.

\begin{figure}[htbp]
\begin{center}
\includegraphics[width=1\columnwidth]{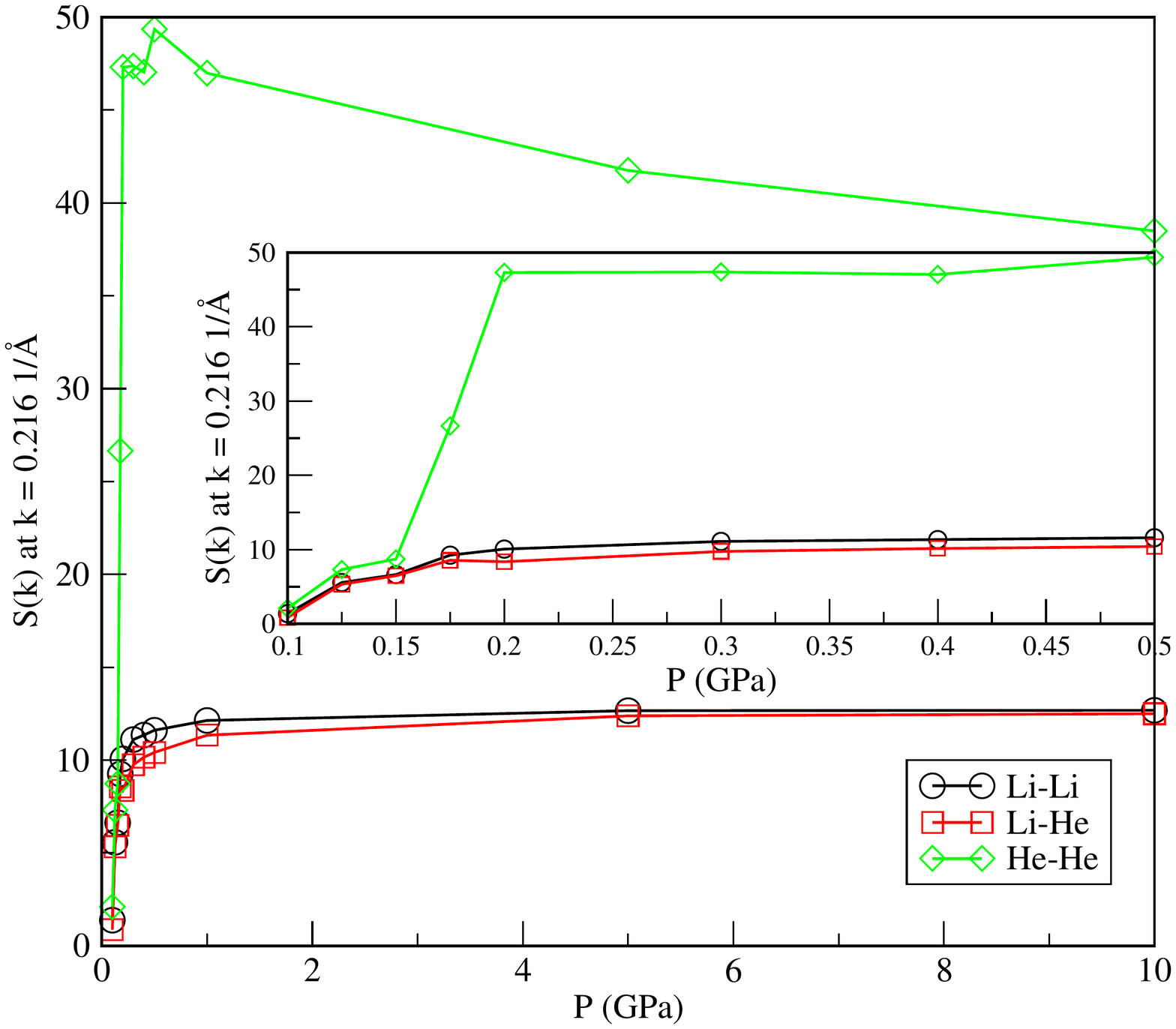}
\caption{\label{fig7} Static structure factors computed at very low momentum ($k = 0.216 1/\AA$) 
a function of the pressure.  Li-Li (black circles);  Li-He (red squares); He-He (green diamonds).}
\end{center}
\end{figure}

In order to further characterize the phase separation, we are also reporting static structure factors $S(k)$ 
computed from the RDF (Eq.~\ref{eq4}) for low momenta ($k = 0.216 ~\AA^{-1}$) as a function of the pressure:

\begin{equation}
S(k) = 1 + 4\pi\rho_{0} \int_{0}^{\infty} {\rm d}r\;r(g_{\alpha, \beta}(r)-1)\sin kr,
\label{eq4}
\end{equation}
where $g_{\alpha, \beta}(r)$ is given by Eq.~\ref{eq4},  and $\rho_0$ is the average density of each species 
($\alpha, \beta$).  The results are shown in Fig.~\ref{fig7}.  The change in the slope of $S(k)$ is particularly 
sharp in the He-He case, around the crossover pressure of 0.175 GPa, differently of the cases of Li-Li and 
Li-He.  It has recently reported for the Ising model~\cite{wang2016discovering} that the sensibility of changes
in slope of static structure factors may be a clear indication of a possible phase transition.  The precise 
quantification of such a phase transition is currently evaluated in our lab,  although it is out of the scope of 
the present work. 

\begin{figure}[htbp]
\begin{center}
\includegraphics[width=1.\columnwidth]{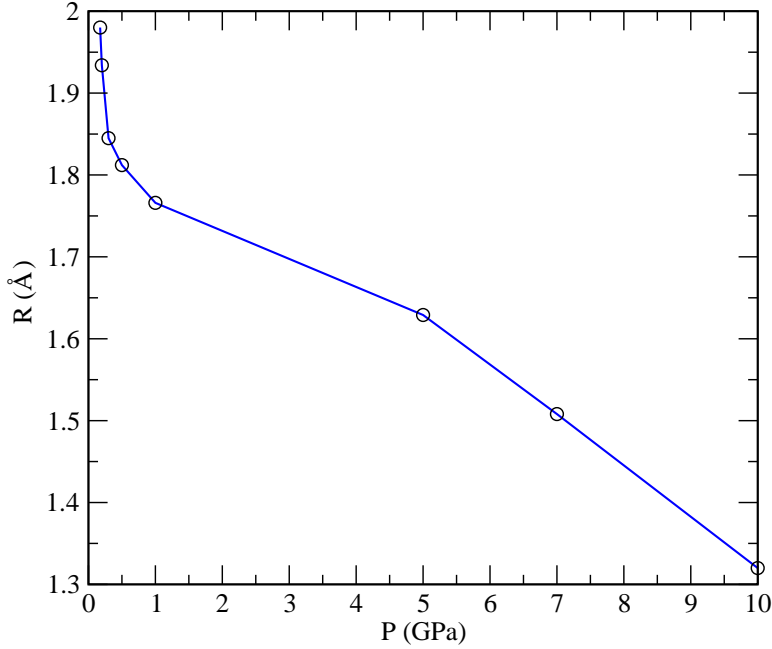}
\caption{\label{fig8} Radii of helium droplets for the pressure range 0.175-10 GPa. }
\end{center}
\end{figure}

The radii of helium droplets formed in our simulations are reported in 
Table~\ref{tab2} and represented in Fig.~\ref{fig8} while the specific size 
depends on the amount of particles in the simulation. We considered only
the reference case of 40 Helium and 960 lithium atoms. One observes that 
the radius is largest at low pressures thus decreasing as the pressure increases. 
At pressures below 1 GPa, we can fit an exponential law: $R = 1.965 e^{-0.12 P}$ 
($P$ in GPa) whereas in the range above 1 GPa, the best fit is linear: 
$R = 1.84-0.05 P$. This indicates a qualitatively different behaviour for $R$, 
strongly dependent on the pressure.

\subsection{Atomic diffusion}
\label{diff}

\begin{figure}[htbp]
\begin{center}
\includegraphics[width=1\columnwidth]{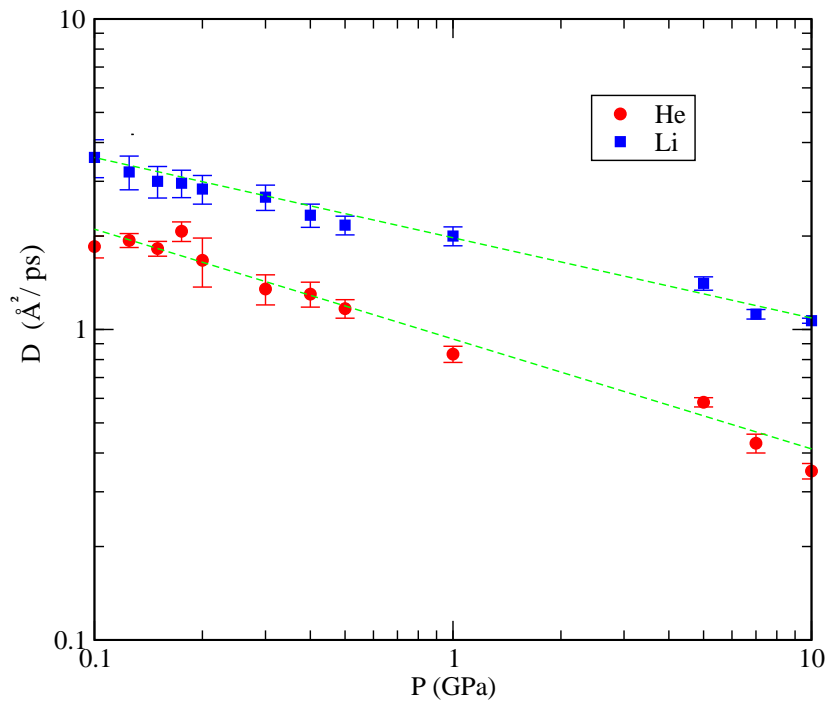}
\caption{\label{fig9} Diffusion coefficients of lithium (blue squares) and helium (red circles) at
  843~K as a function of the of the pressure on a double-logarithmic scale.  Green straight lines
are a guide to the eye.}
\end{center}
\end{figure}

Another experimentally relevant quantity is the diffusion coefficient. Using the 
molecular dynamics method, the mean square displacement (MSD) for both helium 
and lithium was separately obtained. The value of the diffusion coefficient $D$ 
was then computed from the slope of the steady-state MSD,  using Einstein's formula

\begin{equation}
D = \frac{1}{6}\;\lim_{t\rightarrow \infty}\;\frac{{\rm d}}{{\rm d}t}
\langle|{\bf r}(t)- {\bf r}(0)|^2\rangle,
\label{eq5}
\end{equation}
where ${\bf r}$ stands for the coordinate of each species (lithium, helium).  
The coefficients at all simulated states are reported in Table ~\ref{tab3}.
Canales et al. \cite{canales1994computer} obtained a value for the diffusion coefficient of
pure lithium at 843~K (around 1~GPa) of 2.47~\AA$^2$/ps, whereas Jayaram et 
al.~\cite{jayaram2001calculation} reported 0.8~\AA$^2$/ps at 500~K.  Our result at 843~K
is of $D$ = ~2.0~\AA$^2$/ps,  indicates that the lithium diffusion coefficient does not change
significantly from its value in the absence of helium. This is not surprising considering the 
low concentration of helium atoms examined.  It is also worth noticing that the
experimental value of 45~\AA$^2$/ps at 523~K,  reported by Nieto et al.~\cite{nieto2003helium} 
for helium injected onto the surface of a stream of flowing lithium corresponds
to a system out of equilibrium, which is a different situation from the one analyzed here.
This can explain the large difference of two orders of magnitude when compared to
our result,  0.833 ~\AA$^2$/ps at 843~K and 1~GPa (see Table ~\ref{tab3}).  Experiments 
in other similar systems might provide a more suitable reference to compare our results to.  
Figure~\ref{fig9} shows the values of $D$ obtained in our simulations as a function of 
the pressure.  As it can be seen,  the dependence of $D$ on $P$ is approximately linear, 
showing slower diffusion at pressures above 1~GPa.

\subsection{Atomic spectroscopy}
\label{spec}

Experimental infrared spectra are usually obtained through the absorption coefficient 
$\alpha(\omega)$ or the imaginary part of the frequency-dependent dielectric
constant\cite{mcquarrie2000statistical}. These properties are directly related to 
the absorption lineshape $I(\omega)$, which can also be obtained from molecular dynamics
simulations\cite{marti1993computer,marti1994dielectric,praprotnik2005molecular}
in certain situations.  In most cases the physically relevant property to be
computed is the so-called atomic spectral density $S_{i}(\omega)$,  defined as:

\begin{equation}
S_{i}(\omega) = \int_{0}^{\infty}{\rm d}t\;\langle\vec{v}_{i}(t)\vec{v}_{i}(0)\rangle\,\cos(\omega t)\,
\label{eq6}
\end{equation}
where $\vec{v}_{i}(t)$ is the velocity of the $i-th$ atom at time $t$, while the 
brackets $\langle\cdots\rangle$ denote an equilibrium ensemble averaging.
In our case we have obtained the spectral density of
each atomic species separately.  Generally speaking, classical molecular dynamics 
simulations are not able to fully reproduce experimental absorption coefficients, these 
being quantum properties. However they can be used to locate the position of the
spectral bands since in the harmonic (oscillator) approximation,  classical and quantum 
ground state frequencies are equal.

\begin{figure}[htbp]
\begin{center}
\includegraphics[width=1\columnwidth]{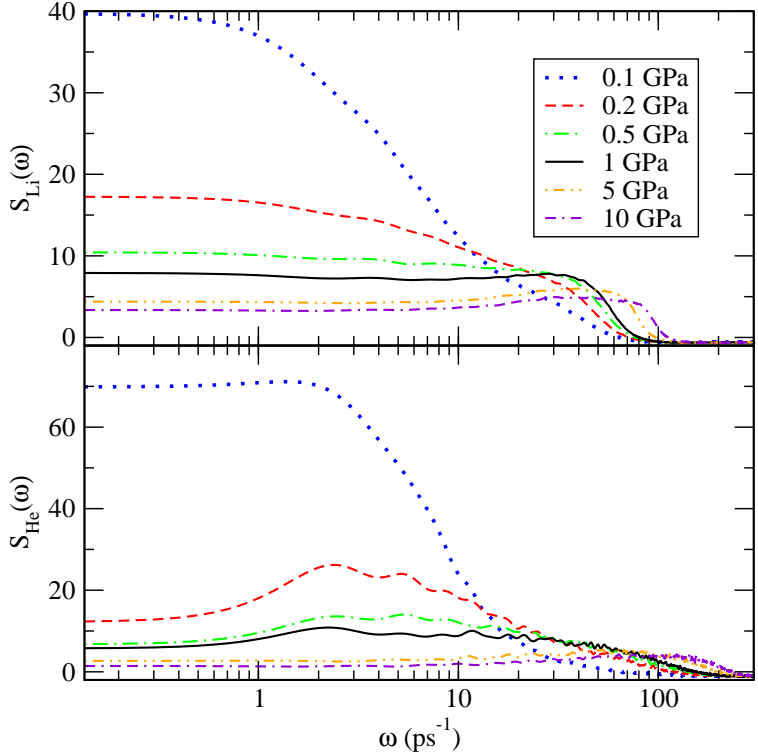}
\caption{\label{fig10} Spectral densities of lithium (top) and helium (bottom)
at 843~K as a function of pressure.  Lines: 0.1 GPa (dotted blue); 0.2 GPa 
(dashed red); 0.5 GPa (dot-dashed green); 1 GPa (black); 5 GPa 
(dot-dot-dashed orange); 10 GPa (dash-dash-dotted violet).}
\end{center}
\end{figure}

The power spectrum describes the main vibrational modes of a molecular
system, including low frequencies below 100~ps$^{-1}$, associated with
translational and rotational modes, and high frequencies of stretching
and bending vibrations around and above 500~ps$^{-1}$. The power
spectra were obtained for the  velocity autocorrelation functions (VACF)
of lithium and helium atoms and are shown in Fig.~\ref{fig10}.  We find
that lithium atoms have a tendency to oscillate at frequencies between 30 and 
85~ps$^{-1}$,  whereas the vibrational frequency for helium atoms is between 2 and
100~ps$^{-1}$, approximately. The fact that these peaks are found at low frequencies 
is consistent with a picture where the atoms in our system can only
present translational vibration modes \cite{padro2004response}.

\section{Concluding Remarks}
\label{concl}

In this work we have analysed the structure and dynamics
of lithium-helium mixtures with a very low He concentration as a first step 
towards the simulation of the typical environmental conditions in the BB of a 
fusion power plant.  We perform ab-initio simulation of the 
lithium-helium mixture using Monte Carlo and molecular dynamics methods, which 
yield the same predictions at equilibrium.  Monte Carlo method is more efficient 
for the calculation of thermodynamic quantities and we employ it for the estimation of
the total energy and pressure,  along with some of its structural properties as the pair 
distribution functions.  From molecular dynamics we have obtained,  in addition,  
time-dependent quantities such as the diffusion coefficients, velocity autocorrelation 
functions for the atoms in the mixture, as well as power spectra of the latter.

In our simulations, helium atoms are miscible in the lithium bath at low pressures whereas, 
for pressures above some critical value, the mixture has the tendency to phase separate 
and split into pure lithium and pure helium phases (formation of helium droplets in the 
lithium background). This tendency is in overall contradiction with
Henry’s law, which has been observed experimentally to be met by
liquid metals\cite{slotnick1965solubility}.

The simulations reported in this work are a first step towards the
understanding of the phenomenon of helium nucleation in liquid
lithum. Although our results do not agree with the behavior predicted
by CNT, we have shown that the phenomenon of helium nucleation in
liquid lithium at high temperatures and pressures can be captured by
MD at its inception; independently of the initial homogeneous
disposition of atoms in the system, our simulation shows the formation
of helium drops systematically if the same environmental conditions are
met. Dynamical properties of the mixture such as diffusion coefficients
of lithium and helium are very be well reproduced, in overall good
agreement with experimental data available. Better potential 
functions are to be tested in future studies which fit within the expected Henry’s 
law behavior. Future studies would likely involve the calculation of surface tensions 
of the droplets and the analysis of the nucleation phenomenon on lithium-lead mixtures.

\begin{acknowledgments}
We thank L.A.Sedano and A.Awad for fruitful discussions. J.M and
L.B. acknowledge financial support from the Generalitat de Catalunya (project 
"FusionCAT", number J-02603). J.M. thanks the Spanish Ministry of Science,
Innovation and Universities (project number PGC2018-099277-B-C21, funds 
MCIU/AEI/FEDER, UE). G.E.A. and F.M. acknowledge financial support from the 
Spanish MINECO (FIS2017-84114-C2-1-P), and from the Secretaria d'Universitats 
i Recerca del Departament d'Empresa i Coneixement de la Generalitat de Catalunya 
within the ERDF Operational Program of Catalunya (project QuantumCat, Ref.~001-P-001644).
\end{acknowledgments}

\begin{table}
\begin{center}
\caption{\label{tab1} Average internal energies ($U$), pressures ($P$)
 and temperatures ($T$) for the simulated setups. All MC simulations
 considered $10^8$ sampling moves and all MD simulations were of
 total length 200 ps.}
\begin{tabular}{c|c|c|c}  \hline
Method & $U$(K) & P (GPa) & T(K) \\  \hline
&  71.1 &  0.104 & 843 \\
&  -187.0 &  0.122 & 843 \\
&  -430.3 &  0.146 & 843 \\
&  -553.3 &  0.178 & 843 \\
MC &  -700.0 &  0.208 & 843 \\
&  -1183.7 &  0.500 & 843 \\
&  -1398.6 &  1.005 & 843 \\
& -1197.6 &  5.002 & 843 \\
& -261.9 &  9.994 & 843 \\  \hline
&  150.8 &  0.105 & 842.3  \\
& -214.2 &  0.128 & 842.1 \\
& -380.4 &  0.152 & 842.1 \\
& -507.6 &  0.177 & 842.1 \\
MD &  -720.8 &  0.202 & 841.9 \\
& -1182.3 &  0.501 & 841.7 \\
& -1405.2 &  0.999 & 841.6 \\
& -1199.1 &  4.999 & 841.0 \\
& -262.8 &  9.998 & 840.3 \\ \hline
\end{tabular}
\end{center}
\end{table}

\begin{table}
\begin{center}
\caption{\label{tab2} Radii of helium drops at 843 K as a function of the pressure.}
\begin{tabular}{c|c}  \hline
Pressure (GPa) & $R_{He}$ (\AA) \\  \hline
0.175 & 1.980 \\
0.2 & 1.934 \\
0.3 & 1.845  \\
0.5 & 1.812 \\
1 & 1.847 \\
5 & 1.629 \\
7 & 1.508  \\
10 & 1.320 \\ \hline
\end{tabular}
\end{center}
\end{table}

\begin{table}
\begin{center}
\caption{\label{tab3} Diffusion coefficients of lithium and helium at 843 K as a function 
of the pressure.}
\begin{tabular}{c|c|c}  \hline
Pressure (GPa) & $D_{He}$ (\AA$^2$/ps)  & $D_{Li}$ (\AA$^2$/ps) \\  \hline
0.1 & 1.850 & 3.583 \\
0.125 & 1.936 & 3.217 \\
0.15 & 1.822 & 3.010 \\
0.175 & 2.071 & 2.958 \\
0.2 & 1.670 & 2.833 \\
0.3 & 1.350 & 2.667 \\
0.4 & 1.300 & 2.333 \\ 
0.5 & 1.167 & 2.167 \\
1 & 0.833 & 2.006 \\
5 & 0.583 & 1.408 \\
7 & 0.430 & 1.120 \\
10 & 0.350 & 1.067 \\ \hline
\end{tabular}
\end{center}
\end{table}

\bibliography{references.bib}

\end{document}